\pdfoutput=1
\documentclass[11pt,twoside,a4paper,cmspaper,final,collab]{cms-tdr}
\def\svnVersion{a759a57}\def\svnDate{2025/03/21}\def\cmsCernNoTag{CERN-EP-2023-292}\def\cmsCernDate{\today}\def\cmsMessage{Published in Reports on Progress in Physics as \href{http://dx.doi.org/10.1088/1361-6633/ad4b9b}{\doi{10.1088/1361-6633/ad4b9b}.}}
\begin{document}\cmsNoteHeader{HIN-23-003}

\newcommand{\etsumhf}{\ensuremath{E_{\text{T, sum}}^{\mathrm{HF}}}\xspace}
\newcommand{\cssquare}{\ensuremath{c^2_{\mathrm{s}}}\xspace}
\newcommand{\cs} {\ensuremath{c_{\mathrm{s}}}\xspace}
\newcommand{\nch}{\ensuremath{N_{\text{ch}}}\xspace}
\newcommand{\nchnorm}{\ensuremath{N_{\text{ch}}^{\text{norm}}}\xspace}
\newcommand{\nchknee}{\ensuremath{\overline{N^{\text{knee}}_{\text{ch}}}}\xspace}
\newcommand{\ptave}{\ensuremath{\langle {\pt} \rangle}\xspace}
\newcommand{\ptavenorm}{\ensuremath{\langle {\pt} \rangle^{\text{norm}}}\xspace}
\newcommand{\PbPb}{\ensuremath{\text{PbPb}}\xspace}

\cmsNoteHeader{HIN-23-003}

\title{Extracting the speed of sound in quark-gluon plasma with ultrarelativistic lead-lead collisions at the LHC}
\titlerunning{Speed of sound in quark-gluon plasma}
\date{\today}

\abstract{
Ultrarelativistic nuclear collisions create a strongly interacting state of hot and dense quark-gluon matter that exhibits a remarkable collective flow behavior with minimal viscous dissipation. To gain deeper insights into its intrinsic nature and fundamental degrees of freedom, we determine the speed of sound in an extended volume of quark-gluon plasma using lead-lead ({\PbPb}) collisions at a center-of-mass energy per nucleon pair of 5.02\TeV. The data were recorded by the CMS experiment at the CERN LHC and correspond to an integrated luminosity of 0.607\nbinv. The measurement is performed by studying the multiplicity dependence of the average transverse  momentum of charged particles emitted in head-on \PbPb collisions. Our findings reveal that the speed of sound in this matter is nearly half the speed of light, with a squared value of $0.241\pm0.002\stat\pm0.016\syst$ in natural units. The effective medium temperature, estimated using the mean transverse momentum, is $219\pm8\syst\MeV$. The measured squared speed of sound at this temperature aligns precisely with predictions from lattice quantum chromodynamic (QCD) calculations. This result provides a stringent constraint on the equation of state of the created medium and direct evidence  for a deconfined QCD phase being attained in relativistic nuclear collisions.
}

\hypersetup{
pdfauthor={CMS Collaboration},
pdftitle={Extracting the speed of sound in quark-gluon plasma with ultrarelativistic lead-lead collisions at the LHC},
pdfsubject={CMS},
pdfkeywords={CMS, quark-gluon plasma, speed of sound, ultra-central, QCD equation of state}}

\maketitle 

\section{Introduction}

When heavy atomic nuclei collide at relativistic speeds, a transformation occurs, giving 
rise to an exotic state of matter with a temperature above several trillion kelvin and 
known as the quark-gluon plasma (QGP)~\cite{Shuryak:1977ut,Collins:1974ky,Cabibbo:1975ig,Freedman:1976ub}. 
In this realm of extreme temperatures, quarks and 
gluons break free from their confined existence inside hadrons, traversing long distances (\eg, several \unit{fm}) compared to the size of individual nucleons.
The emergence of the QGP represents a fundamental prediction of quantum chromodynamics (QCD)~\cite{Hwa:2004yg,HotQCD:2014kol}, 
the theory that elucidates the nature of the strong force.
More remarkably, this strongly interacting QGP matter is found to exhibit the characteristics of 
an almost ``perfect liquid'' with little frictional momentum dissipation~\cite{Adams:2005dq,Adcox:2004mh,Arsene:2004fa,Back:2004je}. 
Its collective dynamics and macroscopic properties are well described by the principles 
of nearly ideal relativistic hydrodynamics.

The equation of state (EoS) reveals the underlying fundamental degrees of freedom of 
a substance and is an invaluable tool to infer how the substance will respond to changes in its energy density.
In fluid-like environments, the study of sound modes arising from 
longitudinal compression provides a means to determine the corresponding speed of sound, denoted as \cs.
This parameter, whose square is defined as the rate of pressure $P$ change in response to 
variations in energy density $\varepsilon$, $\cssquare =\ddinline{P}{\varepsilon}$~\cite{Heinz:2013th}, 
plays a pivotal role in characterizing the nature of the medium under investigation and in constraining models of corresponding EoS.
The exploration of the sound wave propagation in strongly 
correlated systems, ranging from neutron stars to ultracold atomic gases~\cite{2007PhRvL..98q0401J,2020Sci...370.1222P}, 
has garnered significant interest in recent years.
Various methodologies have been proposed to experimentally extract the speed of sound in a QGP fluid
~\cite{Campanini:2011bj,Pratt:2015zsa,Gardim:2019xjs,Gardim:2019brr,Sorensen:2021zme}, offering a direct means to 
constrain the QCD EoS. 
Notably, constraints on the speed of sound in hot QCD matter have been inferred through 
a comparison of relativistic nuclear collision data with theoretical models within a Bayesian framework~\cite{Pratt:2015zsa}.
Recently, an effort to directly extract \cssquare in the QGP phase was made by establishing a connection to 
an effective static, uniform fluid system~\cite{Gardim:2019xjs}.
That work was based on only two independent measurements of the charged-particle multiplicity 
density and mean transverse momentum ({\pt}) in lead-lead ({\PbPb}) collision data from the 
ALICE experiment at center-of-mass energies per nucleon pair $\sqrtsNN=2.76$ and $5.02\TeV$, and
yielded a value of $\cssquare=0.24\pm0.04$ in natural units at a temperature of $222\pm9$\MeV. 
This result is in line with lattice QCD predictions, albeit subject to significant experimental uncertainties.

\begin{figure}[htb]
\centering
\includegraphics[width=\linewidth]{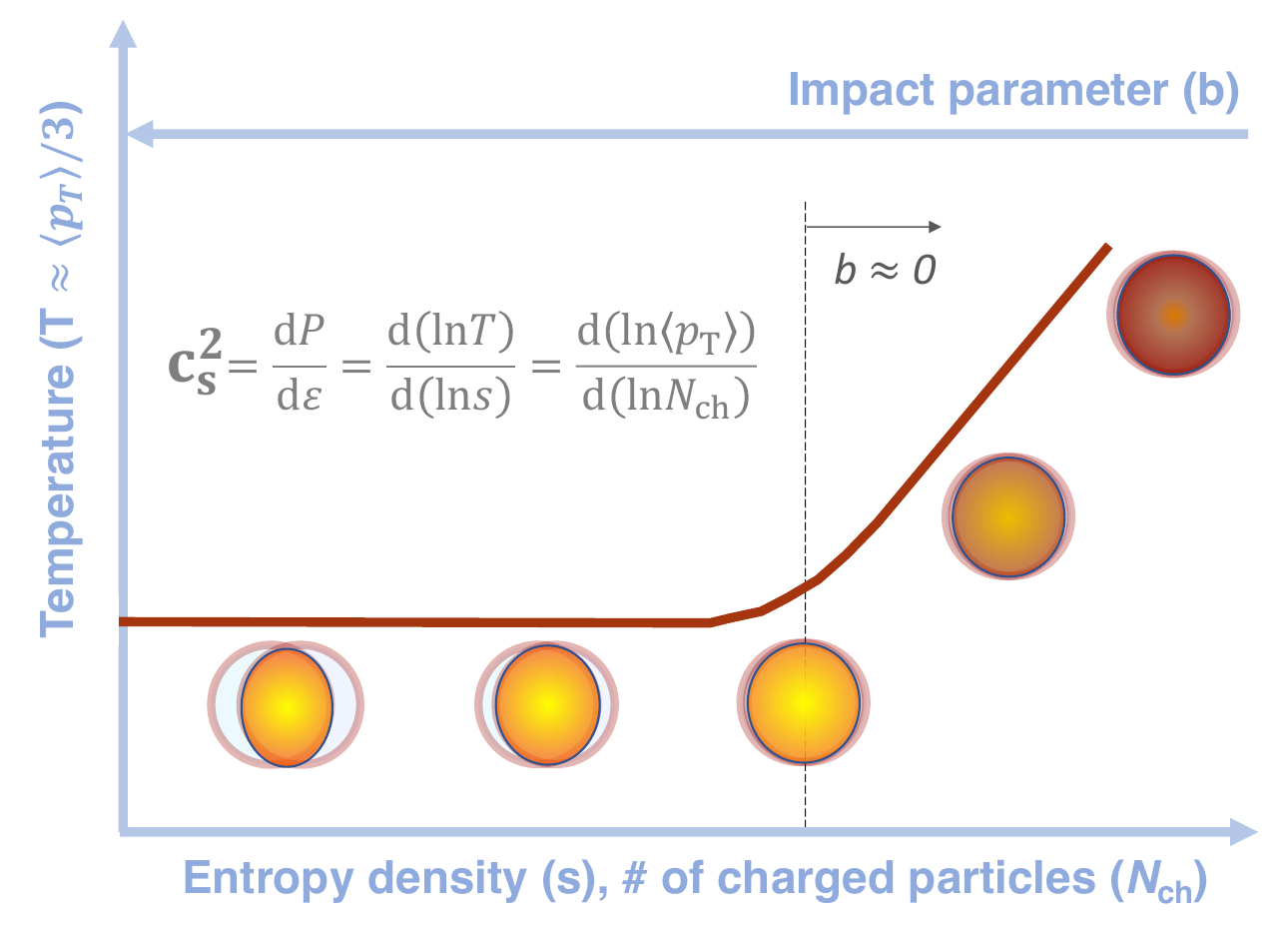}
  \caption{ \label{fig:cartoon}  Conceptual representation of temperature vs. entropy density from mid-central to ultra-central heavy ion collisions. }
\end{figure}

To increase the precision by which the speed of sound can be determined,
a new hydrodynamic probe was later proposed in Ref.~\cite{Gardim:2019brr} utilizing the multiplicity dependence of mean \pt measurements at a fixed $\sqrtsNN$.
This innovative technique makes use of ``ultra-central'' collisions in which 
the ions overlap almost entirely, \ie, collide at a very small impact parameter ($b$).
A conceptual representation of this probe is illustrated in Fig.~\ref{fig:cartoon}. 
The impact parameter of a heavy ion collision determines the size of the nuclear overlap region (system size), which is strongly correlated
with the energy and entropy deposited in the initial state and the number of emitted charged particles in the final state (``multiplicity'', \nch). 
As the impact parameter decreases and collisions become increasingly central,
both the system size and deposited energy increase, while maintaining a nearly constant initial energy density and temperature.
However, this trend reaches its limit when $b\to0$. 
In this case, the initial system size is limited by the sizes of the participating nuclei. 
For symmetric \PbPb collisions, this would be the size of a Pb nucleus. 
More energy and entropy can still be deposited into the fixed volume through fluctuations in the number of interacting partons.
By examining the response of the temperature $T$ to the increasing 
entropy density $s$ at $b\sim0$, the speed of sound can be extracted based on fundamental thermodynamic laws, 
\begin{equation}
\label{eq:sound2}
\cssquare = \frac{\rd P}{\rd\varepsilon}=\frac{s\rd T}{T\rd s}=\frac{\rd\ptave/\ptave}{\rd\nch/\nch}. 
\end{equation}
\noindent Here, in terms of experimental observables, $s$ is directly proportional to \nch, 
while the temperature $T$ relates to the average transverse momentum ({\ptave}) of emitted particles 
with respect to the beam axis~\cite{Gardim:2019xjs}. Full hydrodynamic simulations, 
such as those made possible using the \textsc{Trajectum} model~\cite{Nijs:2021clz}, 
have verified the above relationship, although
there are features that are not captured, as will be discussed later.
As the \cssquare value depends only on the relative variation in $\ptave$ and $\nch$,
any global changes to the observables, such as an increase in the system entropy through 
hadronic resonance decays~\cite{Qiu:2012tm}, will not affect the result.

In this paper, we present a precise determination of the speed of sound in QGP
using ultra-central \PbPb collision data at $\sqrtsNN=5.02\TeV$, 
collected in 2018 by the CMS experiment at the CERN LHC.
By achieving a level of precision of several percent, 
comparable to theoretical uncertainties, our results serve as a robust benchmark for comparison with 
hydrodynamic simulations and lattice QCD calculations of the EoS. 
These comparisons provide the most stringent and direct constraints on the degrees of freedom 
attained by the medium created in these collisions. Tabulated results are provided in the HEPData record for this analysis~\cite{hepdata}.

\section{The CMS detector}

The CMS apparatus~\cite{CMS:2008xjf} is a multipurpose, nearly hermetic detector, designed to trigger on~\cite{CMS:2020cmk,Khachatryan:2016bia} and identify electrons, muons, photons, and hadrons~\cite{CMS:2020uim,CMS:2018rym,CMS:2014pgm}. The initial triggering is done with the level-1 system, which uses customized hardware to make the rapid online decision whether or not to accept an event and deliver it to the second system, the High Level Trigger (HLT). The HLT uses a large CPU farm to perform optimized online event reconstruction and characterize an event. A global ``particle-flow'' algorithm~\cite{CMS:2017yfk} aims to reconstruct all individual particles in an event, combining information provided by the all-silicon pixel and strip tracker, and by the crystal electromagnetic and brass-scintillator hadron calorimeters,
operating inside a 3.8\unit{T} superconducting solenoid, with data from the gas-ionization muon detectors embedded in the flux-return yoke outside the solenoid. 
Hadron forward (HF) calorimeters~\cite{Bayatian:2006jz}, made of steel and quartz fibers, extend the pseudorapidity ($\eta=-\ln(\tan(\theta/2))$, where the polar 
angle $\theta$ is defined relative to the counterclockwise beam) coverage provided by the barrel and endcap detectors.
Two zero-degree calorimeters (ZDCs)~\cite{Suranyi:2021ssd}, made of quartz-fibers and plates embedded in tungsten absorbers, are used to detect neutrons from nuclear dissociation events.

\section{Data samples, event reconstruction and selection}

The data analyzed, before applying the selection described below, consist of 4.27\ten{9} minimum 
bias events, corresponding to an integrated luminosity of 0.607\nbinv. The minimum bias events are triggered 
by requiring total energy signals above readout thresholds, which are in the range $6\text{--}12\GeV$, 
on both sides of the HF calorimeters~\cite{Khachatryan:2016bia}.
Beam-gas interactions and nonhadronic collisions are rejected by requiring the shapes of the
clusters in the pixel tracker to be compatible with those expected from particles
produced by a \PbPb collision~\cite{CMS:2016xef}.
The events are also required to have at least one reconstructed primary vertex associated with two or more tracks
within a distance of 15\cm from the nominal interaction point along the beam axis.
The primary vertex is selected as the one with the highest track multiplicity in the event. 
Events with concurrent interactions per bunch crossing contribute to about 0.5\% of the full data sample and 
are rejected based on the correlation of total energy deposited 
in the HF and ZDC detectors, following the procedure used in Ref.~\cite{CMS:2013bza}. 
The collision centrality in \PbPb events, \ie, the degree of overlap or impact parameter of the two colliding nuclei, is commonly determined by the total transverse energy deposit in both HF calorimeters,  \etsumhf~\cite{CMS:2016xef}. As the main focus of this work is on collisions at small impact parameters, 
we analyzed only the 10\% of \PbPb events that had the largest \etsumhf.
This class contains the ultra-central collision events of interest.

To ease the computational load for high-multiplicity central \PbPb collisions,
track reconstruction for \PbPb events is done in two iterations.
The first iteration reconstructs tracks from signals
 (``hits'') in the silicon pixel and strip tracker that are compatible with trajectories of particles with $\pt>1.0\GeV$, while
the second iteration reconstructs tracks compatible with trajectories of particles with $0.3<\pt<1.0\GeV$ using solely the pixel detector.
In the analysis, the tracks have the additional selection requirement of $\abs{\eta}<0.5$ for the best tracking performance.
More details on the track reconstruction and selection can be found in Ref.~\cite{CMS:2019nct}.
The tracking efficiency ($\varepsilon_{\text{eff}}$) and misreconstruction rate 
($\varepsilon_{\text{mis}}$) are evaluated using the \HYDJET~\cite{Lokhtin:2008xi} event generator, 
together with a full \GEANTfour~\cite{Agostinelli:2002hh} simulation of the CMS detector response. 
These factors are combined to obtain an overall correction factor, 
$\varepsilon_{\text{trk}}=\varepsilon_{\text{eff}}/(1-\varepsilon_{\text{mis}})$, which
is used to account for detector effects on the total number of reconstructed tracks. The $\varepsilon_{\text{trk}}$ factor is calibrated not only in terms of \pt and $\eta$, but also as a function of
the detector occupancy. The occupancy is estimated by the total number of clusters registered in the 
silicon pixel tracker $N_{\text{pixel}}$, where a weak linear decline of $\varepsilon_{\text{trk}}$ by up to 7\% 
over an increase of $N_{\text{pixel}}$ by 30\% is observed.
In the analysis, each track is assigned a weight of $1/\varepsilon_{\text{trk}}(\eta, \pt, N_{\text{pixel}})$ 
to account for track reconstruction effects.

\section{Measurement method}

The main experimental observable of this analysis is the mean transverse momentum \ptave of 
charged particles in an event as a function of \nch, where \ptave and \nch
are measured within the same $\eta$ and \pt ranges
(otherwise, rapidity-dependent entropy fluctuations would lead to a reduced signal~\cite{Gardim:2019brr}).
Charged particle \pt spectra for $\pt>0.3\GeV$ are measured for events in 50\GeV intervals of \etsumhf from 3400\GeV to 5200\GeV, 
with tracking efficiency and misreconstruction effects corrected.
To avoid any bias in estimating \ptave and \nch, it is necessary to extrapolate the measured \pt spectra to the full \pt range.
The resulting \ptave values (mean of the \pt spectra) from all \etsumhf intervals are then plotted against 
the corresponding \nch values (integral of the \pt spectra) to form the final observable.
The \etsumhf variable essentially serves as a centrality estimator to vary the initial medium entropy density and temperature.
In particular, as the \etsumhf values are obtained in a forward $\eta$ range that does not overlap with the range 
used to measure the corresponding \ptave and \nch values, potential biases are avoided. For example, hard processes originating 
early in the collision tend to fragment into large numbers of high-{\pt} particles, yet these particles may not reflect an increase 
in the entropy and temperature of the QGP medium.

The extrapolation of the \pt spectra to the full \pt range is performed by fitting a Hagedorn function~\cite{Hagedorn:1983wk} 
to the measured \pt spectra over the range of $0.4<\pt<4.5\GeV$ in each \etsumhf interval.
This method is found to provide an excellent description of the data~\cite{ALICE:2018hza} and models (\textsc{Trajectum} and \HYDJET).
The chosen \pt range for the fitting is varied to the evaluate corresponding uncertainties.
The fitted functions are then used to extrapolate the missing portions of the \pt spectra in the low-{\pt} region.

\begin{figure}[htb]
\centering
\includegraphics[width=\linewidth]{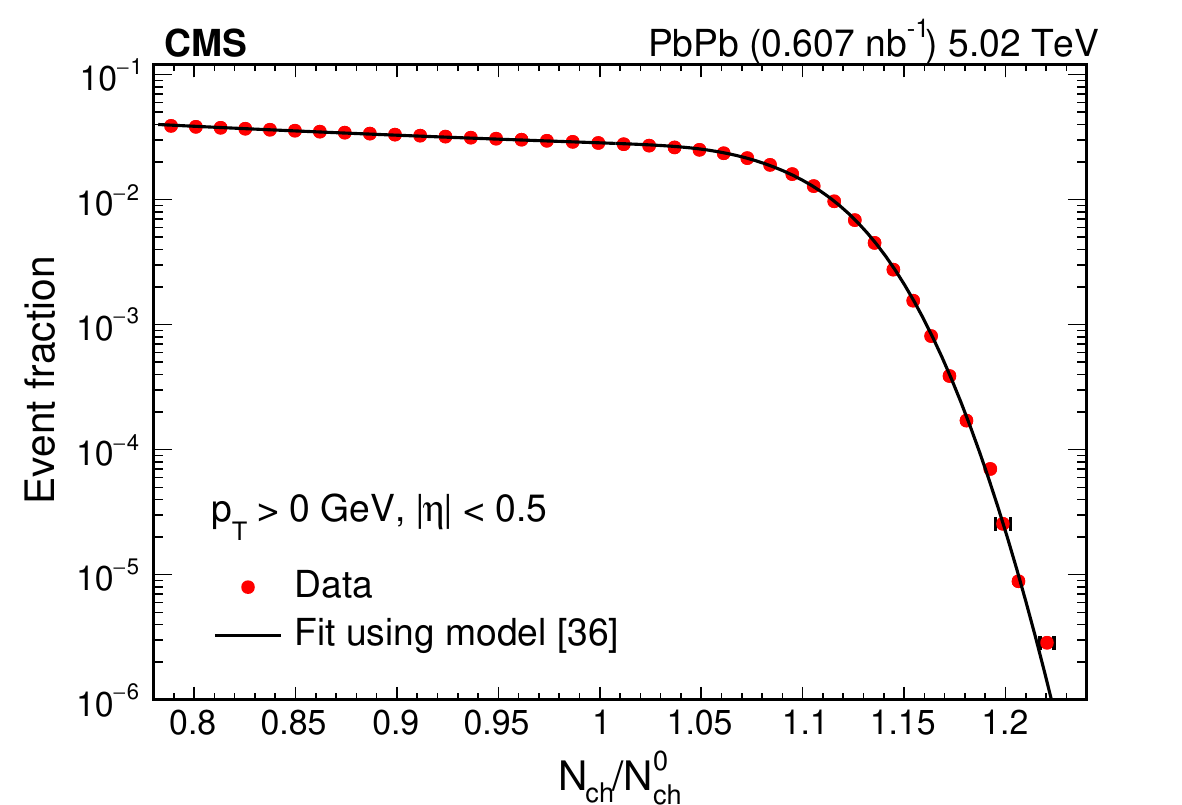}
  \caption{ \label{fig:poisson} The event fraction distribution as a function of the charged-particle multiplicity, \nch, within the kinematic range of $\abs{\eta}<0.5$ and extrapolated to the full \pt range, in \PbPb collisions at $\sqrtsNN=5.02\TeV$. The \nch value is normalized by its value in the 0--5\% centrality class ($\nch^{0}$). The curve represents a fit to the data using the S.~Das \etal model~\cite{Das:2017ned}.}
\end{figure}

As the extraction of the speed of sound mainly depends on the relative variation of \ptave
with respect to \nch (see Eq.~(\ref{eq:sound2})),
normalized quantities, $\ptavenorm =  \ptave/\ptave^{0}$ and $\nchnorm = \nch/\nch^{0}$, are used as the primary observables,
where the $\ptave^{0}$ and $\nch^{0}$ represent the mean transverse momentum and 
charged-particle multiplicity in a reference event class. 
Here, the centrality range chosen for the reference event class only needs to be close to that used for the speed of sound determination, and
5\% most central events (as determined by \etsumhf and denoted ``0--5\%'') is used. By normalizing both \ptave and \nch by their values
in the reference event class, most of the systematic uncertainties can be minimized. 
The $\ptave^{0}$ and $\nch^{0}$ values obtained are found to be  
in good agreement with the ALICE results in the 0--5\% centrality range~\cite{ALICE:2015juo,ALICE:2018hza}.
Figure~\ref{fig:poisson} shows the event fraction distribution as a function of the normalized multiplicity.

To extract the speed of sound, the expression that describes \ptavenorm as a function of \nchnorm 
is taken from Ref.~\cite{Gardim:2019brr}, as
\begin{equation}
\label{eq:sound4}
\ptavenorm = \left(\frac{\nchnorm}{\langle \nchknee | \nchnorm \rangle}\right)^{\cssquare},
\end{equation}
\noindent where, 
\begin{equation}
\label{modeli}
\langle \nchknee | \nchnorm \rangle=
\nchnorm-\sigma\sqrt{\frac{2}{\pi}}
\frac{\exp\left(-\frac{(\nchnorm -\nchknee)^2}{2\sigma^2}\right)}{\erfc\left(\frac{\nchnorm -\nchknee}{\sqrt{2}\sigma}\right)}.
\end{equation}
Here, \nchknee and $\sigma$ represent the mean and root-mean-square width of the charged-particle multiplicity distribution at $b=0$,
normalized by $\nch^{0}$.  
In Fig.~\ref{fig:poisson}, the \nchknee value corresponds to the vicinity of the location beyond which the knee-shaped distribution starts rapidly falling.
For the region of $ \nchnorm<\nchknee$,  the $\langle \nchknee | \nchnorm \rangle$ variable approximately reduces to $\nchnorm$, so Eq.~(\ref{eq:sound4}) yields a value of unity. 
For the region of $\nchnorm>\nchknee$, the $\langle \nchknee | \nchnorm \rangle$ variable saturates at \nchknee for sufficiently large \nchnorm. 
In this limit, Eq.~(\ref{eq:sound4}) becomes a simple power function, with \cssquare being the power of the function. 
The parameters \nchknee and $\sigma$ can be constrained by fitting the measured multiplicity distribution 
using the procedure described in Ref.~\cite{Das:2017ned}. The multiplicity distribution
at fixed values of $b$ is modeled using a Gaussian function. Integrating over $b$ gives a minimum bias 
multiplicity distribution which can be fitted to data. As shown in Fig.~\ref{fig:poisson}, this fit provides 
a good description of the data. The results of this fit can be used to estimate the Gaussian mean and 
width at $b=0$, yielding $\nchknee = 1.11$ and $\sigma = 0.0272$ with negligible uncertainties.
Using the extracted $\nchknee$ and $\sigma$ values, a fit to the measured $\ptavenorm$ 
as a function of $\nchnorm$ is performed using Eq.~(\ref{eq:sound4}), thereby extracting the speed of sound.
In practice, we limit the fit to the very high-multiplicity region of $\nchnorm > 1.14$, as will be discussed in detail later.

The dominant sources of systematic uncertainties for the measured $\ptavenorm$ and $\nchnorm$ values 
originate from the tracking correction and the extrapolation to the full \pt range. 
As mentioned earlier, using normalized quantities minimizes the majority of the systematic uncertainties. 
Systematic uncertainties are directly evaluated for the normalized
quantities, as well as for $\ptave^{0}$ and \cssquare. The tracking correction uncertainty is evaluated by varying
the default track selections to a set of looser or tighter values. The maximum deviation with respect to the default results
is taken as a systematic uncertainty, which is found to be ${\pm}0.01\GeV$ in $\ptave^{0}$ and ${\pm}0.002$ in the fitted \cssquare value.
The \pt extrapolation uncertainty is estimated by varying the range of measured spectra fitted by the Hagedorn function to a lower limit of 0.3 or 0.5\GeV and an upper limit of 4 or 5\GeV. The resulting systematic uncertainty is found to be at most ${\pm}0.023\GeV$ for $\ptave^{0}$ and ${\pm}0.012$ for the \cssquare value. 
Systematic uncertainties for \cssquare associated with the choice of the lower fit limit in $\nchnorm$ are estimated by varying the limit from 1.13 to 1.17, 
resulting in an uncertainty of ${\pm}0.010$ in \cssquare.
Total uncertainties are obtained by adding the various sources in quadrature. 
Systematic uncertainties for $\ptavenorm$ are extracted point-by-point as a function of $\nchnorm$.

\section{Results}

\begin{figure}[htb]
\centering
\includegraphics[width=\linewidth]{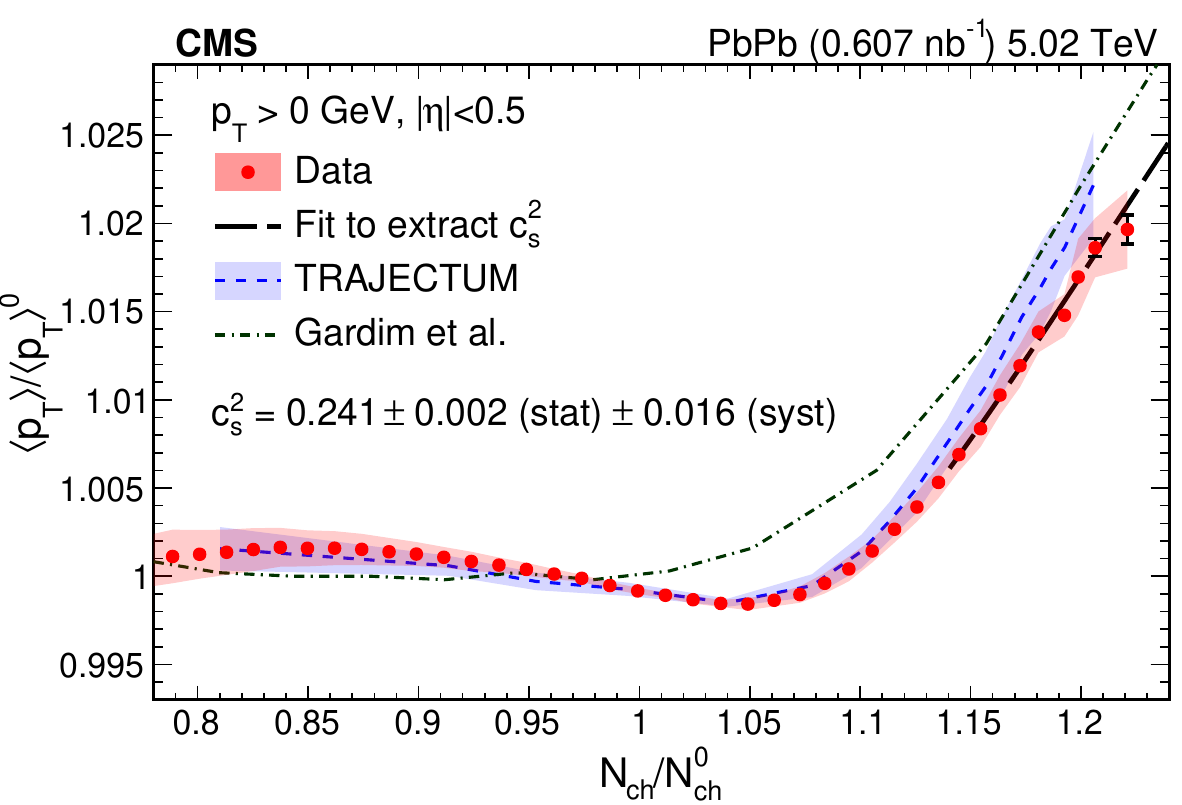}
  \caption{ \label{fig:sosfinal_pt0}  The average transverse momentum of charged particles, \ptave, as a function of
the charged-particle multiplicity, \nch, within the kinematic range of $\abs{\eta}<0.5$ and extrapolated to the full \pt range  in \PbPb collisions at $\sqrtsNN=5.02\TeV$.
Both \ptave and \nch are normalized by their values in the 0--5\% centrality class ($\ptave^{0}$ and $\nch^{0}$).
Bars and the red band correspond to statistical and systematic uncertainties, respectively.
Hydrodynamic simulations from the \textsc{Trajectum} model~\cite{Nijs:2021clz} and the model by Gardim \etal~\cite{Gardim:2019brr} are also shown for comparison. 
The dashed line is a fit to the data using Eq.~(\ref{eq:sound4}) in the range of $\nch/\nch^{0}>1.14$. }
\end{figure}

The observed multiplicity dependence of the average transverse momentum, both normalized by their values in the 0--5\% centrality class,
is presented in Fig.~\ref{fig:sosfinal_pt0}, within the kinematic range of $\abs{\eta}<0.5$ and extrapolated to the full \pt range in central \PbPb events.
Hydrodynamic simulations from the \textsc{Trajectum}~\cite{Nijs:2021clz,Giacalone:2023cet,Nijs:2023bzv} 
and Gardim \etal~\cite{Gardim:2019brr} models are also shown for comparison. 
Both models use an EoS from lattice QCD calculations~\cite{Borsanyi:2010cj}.
The \textsc{Trajectum} model is a computational framework to simulate the full evolution of
heavy ion collisions, which includes the modeling of initial stages, a viscous hydrodynamic
phase with transport coefficients, and a hadronic gas phase. 
Parameters of the \textsc{Trajectum} model are constrained by a global Bayesian analysis of a variety of experimental observables~\cite{Nijs:2021clz},
where the band shown corresponds to uncertainties within the allowed range of \textsc{Trajectum} configuration parameters. 
The model of Gardim \etal~\cite{Gardim:2019brr}, besides the hydrodynamic phase, 
also considers the preequilibrium dynamics and hadronic interactions after thermal freeze-out.  
No uncertainties are evaluated for this model as only a single set of model parameters is used.

The $\ptavenorm$ value first shows a very weak declining trend toward a local minimum around $\nchnorm\sim1.05$. 
At higher multiplicities, corresponding to ultra-central \PbPb events, a steep rise is observed,
which is consistent with the expected increase in temperature with entropy density, as schematically illustrated in Fig.~\ref{fig:cartoon}.
The observed trend, including the minimum around $\nchnorm\sim1.05$,
is qualitatively consistent with the prediction by the \textsc{Trajectum} model. 
A slightly steeper rise at high multiplicities is observed for the \textsc{Trajectum} simulation when compared with the data. 
This suggests that the speed of sound used in the model may be slightly larger than is found in the QGP.
However, this difference is not significant within experimental and theoretical uncertainties.
The model by Gardim \etal also predicts a rise of $\ptavenorm$ at very high multiplicities, with a slope similar to that observed in the data.
However, it shows a flat trend at lower multiplicities instead of the local minimum structure around $\nchnorm\sim1.05$ as seen in the data and the \textsc{Trajectum} model.
The origin of the observed local minimum is not currently understood. 

To directly extract the speed of sound,
the multiplicity dependence of the $\ptavenorm$ data in Fig.~\ref{fig:sosfinal_pt0} is fitted by Eq.~(\ref{eq:sound4}).
Because the observed local minimum is not captured by the simplified model in Eq.~(\ref{eq:sound4}),
the fit is performed only in the high-multiplicity range with $\nchnorm>1.14$.
The final result of the squared speed of sound is found to be  $\cssquare=0.241\pm0.002\stat\pm0.016\syst$ in natural units. 
The same fit is also performed to the prediction from the \textsc{Trajectum} model, resulting in $\cssquare=0.283\pm0.045$, 
where the model uncertainty is again determined within the allowed parameter space
constrained by a global Bayesian analysis~\cite{Nijs:2021clz}.

\begin{figure}[htb]
\centering
\includegraphics[width=\linewidth]{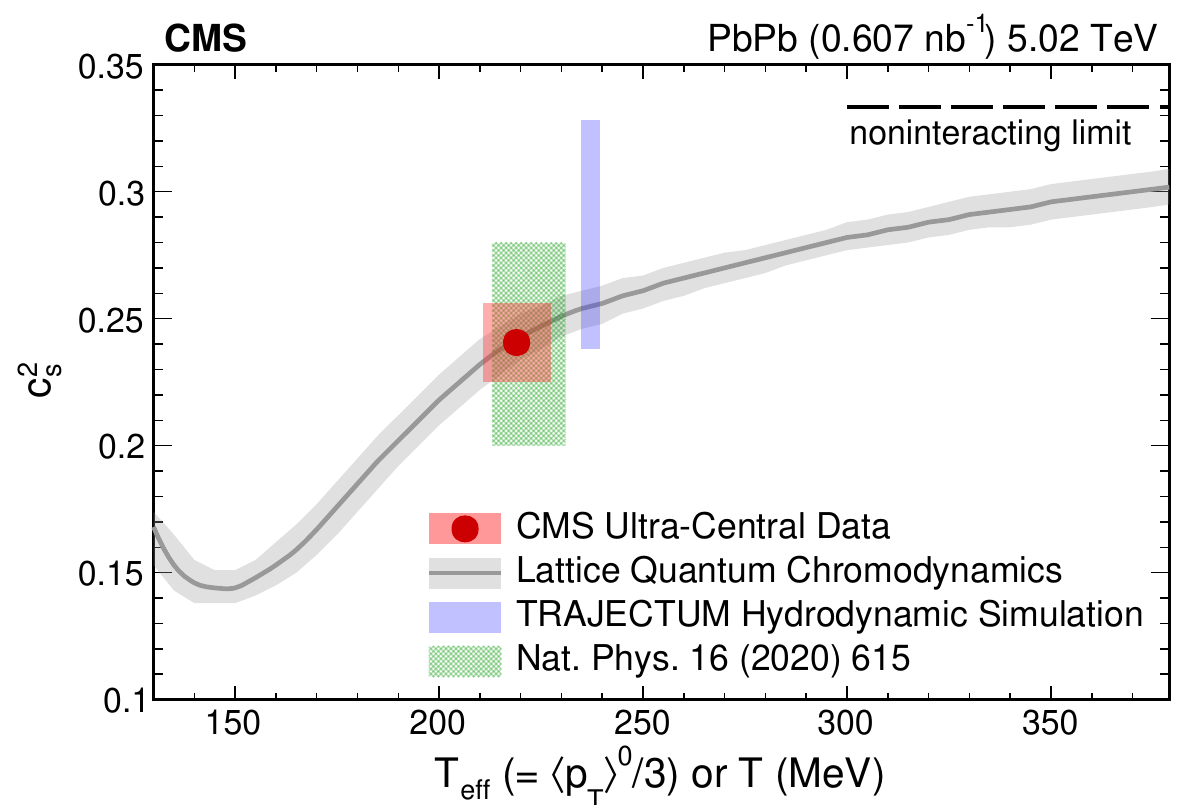}
  \caption{ \label{fig:cs_final} The speed of sound, \cssquare, as a function of the effective temperature, 
  $T_{\text{eff}}$, with the CMS data point obtained from ultra-central \PbPb collision data at $\sqrtsNN=5.02\TeV$.
  The size of the red box indicates systematic uncertainties of \cssquare and $T_{\text{eff}}$, while statistical uncertainties are smaller than the marker size. 
  Values extracted from the \textsc{Trajectum} simulation~\cite{Nijs:2021clz} following the same fitting 
  procedure as the data and from the earlier work~\cite{Gardim:2019xjs} are presented as the other colored boxes. 
  The curve shows the prediction of \cssquare as a function of $T$ from lattice QCD calculations~\cite{HotQCD:2014kol}. 
  The dashed line at the value of $1/3$ corresponds to the upper limit for noninteracting, massless gas (``ideal gas'') systems~\cite{Borsanyi:2010cj}.}
\end{figure}

To constrain the EoS, a simultaneous determination of \cssquare and its corresponding temperature is necessary.
Based on the hydrodynamic simulations discussed in Refs.~\cite{Gardim:2019xjs,Gardim:2019brr}, the 
effective temperature ($T_{\text{eff}}$) of the QGP phase is found to be given 
approximately by $\ptave/3$, with $T_{\text{eff}}=\ptave/3.07$ quoted in 
Ref.~\cite{Gardim:2019xjs} based on a soft equation of state. While the scaling factor relating $T_{\text{eff}}$ to \ptave can depend on specific model assumptions, the theoretical uncertainty in this value is believed to be small compared to the quoted experimental uncertainties, thereby having no impact 
on the main conclusions drawn in this paper.
In essence, $T_{\text{eff}}$ represents the initial temperature that a uniform fluid at rest would have 
if it possessed the same amount of energy and entropy as the QGP fluid does when it reaches 
its freeze-out state, the point at which the quarks become bound into hadrons.
Due to longitudinal expansion and cooling, the $T_{\text{eff}}$ value is generally lower than 
the initial temperature of the QGP fluid. Nevertheless, it still characterizes a temperature in the QGP phase, 
to which the extracted \cssquare value based on the final-state \ptave and \nch corresponds.
Possible effects of shear and bulk viscosity are investigated in Ref.~\cite{Gardim:2019xjs} and found to not impact this framework,
as the shear viscosity increases $\ptave$ by about the same amount that the bulk viscosity decreases it.
The $\ptave^{0}$ value is measured to be $658\pm25\syst\MeV$, leading to a $T_{\text{eff}}$ value for the ultra-central \PbPb data
of $219\pm8\syst\MeV$ (it varies by at most 2\% toward the very end of \nch distribution within the 0--5\% centrality range).
The statistical uncertainty is orders of magnitude smaller than the quoted systematic uncertainties. 

Figure~\ref{fig:cs_final} depicts \cssquare as a function of $T_{\text{eff}}$, 
with the CMS data point obtained from ultra-central \PbPb collision data at $\sqrtsNN=5.02\TeV$.
The results are compared to the \textsc{Trajectum} model, the \cssquare value extracted in Ref.~\cite{Gardim:2019xjs}, and
lattice QCD predictions of the \cssquare value as a function of $T$~\cite{HotQCD:2014kol}.
The new CMS data allow for an unprecedented level of precision in the experimental determination 
of the speed of sound in an extended volume of QGP matter. The results exhibit excellent agreement with the lattice QCD prediction, 
with comparable uncertainties. Thus, our findings provide compelling and 
direct evidence for the formation of a deconfined QCD phase at LHC energies.

\section{Conclusion}

In summary, this study presents a measurement with a new hydrodynamic probe in ultrarelativistic nuclear collisions 
that results in the most precise determination to date of the speed of sound in an extended volume of quark-gluon plasma matter.
By determining the dependence of the average transverse momentum on the total multiplicity for charged particles 
in nearly head-on lead-lead collisions at a center-of-mass energy per nucleon pair of 5.02\TeV, a squared speed of sound of $0.241\pm0.002\stat\pm0.016\syst$ 
in natural units is determined. The effective medium temperature, estimated using the mean transverse momentum, is $219\pm8\syst\MeV$.
The excellent agreement of lattice quantum chromodynamics predictions with the experimental results provides 
strong evidence for the existence of a deconfined phase of matter at extremely high temperatures.

\begin{acknowledgments}
We thank Fernando Gardim, Andre Veiga Giannini, Govert Nijs, Jean-Yves Ollitrault, and Wilke van der Schee for providing us with the model calculations used in Fig.~\ref{fig:sosfinal_pt0}.

We congratulate our colleagues in the CERN accelerator departments for the excellent performance of the LHC and thank the technical and administrative staffs at CERN and at other CMS institutes for their contributions to the success of the CMS effort. In addition, we gratefully acknowledge the computing centers and personnel of the Worldwide LHC Computing Grid and other centers for delivering so effectively the computing infrastructure essential to our analyses. Finally, we acknowledge the enduring support for the construction and operation of the LHC, the CMS detector, and the supporting computing infrastructure provided by the following funding agencies: SC (Armenia), BMBWF and FWF (Austria); FNRS and FWO (Belgium); CNPq, CAPES, FAPERJ, FAPERGS, and FAPESP (Brazil); MES and BNSF (Bulgaria); CERN; CAS, MoST, and NSFC (China); MINCIENCIAS (Colombia); MSES and CSF (Croatia); RIF (Cyprus); SENESCYT (Ecuador); MoER, ERC PUT and ERDF (Estonia); Academy of Finland, MEC, and HIP (Finland); CEA and CNRS/IN2P3 (France); SRNSF (Georgia); BMBF, DFG, and HGF (Germany); GSRI (Greece); NKFIH (Hungary); DAE and DST (India); IPM (Iran); SFI (Ireland); INFN (Italy); MSIP and NRF (Republic of Korea); MES (Latvia); LAS (Lithuania); MOE and UM (Malaysia); BUAP, CINVESTAV, CONACYT, LNS, SEP, and UASLP-FAI (Mexico); MOS (Montenegro); MBIE (New Zealand); PAEC (Pakistan); MES and NSC (Poland); FCT (Portugal); MESTD (Serbia); MCIN/AEI and PCTI (Spain); MOSTR (Sri Lanka); Swiss Funding Agencies (Switzerland); MST (Taipei); MHESI and NSTDA (Thailand); TUBITAK and TENMAK (Turkey); NASU (Ukraine); STFC (United Kingdom); DOE and NSF (USA).

\hyphenation{Rachada-pisek} Individuals have received support from the Marie-Curie program and the European Research Council and Horizon 2020 Grant, contract Nos.\ 675440, 724704, 752730, 758316, 765710, 824093, and COST Action CA16108 (European Union); the Leventis Foundation; the Alfred P.\ Sloan Foundation; the Alexander von Humboldt Foundation; the Science Committee, project no. 22rl-037 (Armenia); the Belgian Federal Science Policy Office; the Fonds pour la Formation \`a la Recherche dans l'Industrie et dans l'Agriculture (FRIA-Belgium); the Agentschap voor Innovatie door Wetenschap en Technologie (IWT-Belgium); the F.R.S.-FNRS and FWO (Belgium) under the ``Excellence of Science -- EOS" -- be.h project n.\ 30820817; the Beijing Municipal Science \& Technology Commission, No. Z191100007219010 and Fundamental Research Funds for the Central Universities (China); the Ministry of Education, Youth and Sports (MEYS) of the Czech Republic; the Shota Rustaveli National Science Foundation, grant FR-22-985 (Georgia); the Deutsche Forschungsgemeinschaft (DFG), under Germany's Excellence Strategy -- EXC 2121 ``Quantum Universe" -- 390833306, and under project number 400140256 - GRK2497; the Hellenic Foundation for Research and Innovation (HFRI), Project Number 2288 (Greece); the Hungarian Academy of Sciences, the New National Excellence Program - \'UNKP, the NKFIH research grants K 124845, K 124850, K 128713, K 128786, K 129058, K 131991, K 133046, K 138136, K 143460, K 143477, 2020-2.2.1-ED-2021-00181, and TKP2021-NKTA-64 (Hungary); the Council of Science and Industrial Research, India; ICSC -- National Research Center for High Performance Computing, Big Data and Quantum Computing, funded by the EU NexGeneration program (Italy); the Latvian Council of Science; the Ministry of Education and Science, project no. 2022/WK/14, and the National Science Center, contracts Opus 2021/41/B/ST2/01369 and 2021/43/B/ST2/01552 (Poland); the Funda\c{c}\~ao para a Ci\^encia e a Tecnologia, grant CEECIND/01334/2018 (Portugal); the National Priorities Research Program by Qatar National Research Fund;  MCIN/AEI/10.13039/501100011033, ERDF ``a way of making Europe", and the Programa Estatal de Fomento de la Investigaci{\'o}n Cient{\'i}fica y T{\'e}cnica de Excelencia Mar\'{\i}a de Maeztu, grant MDM-2017-0765 and Programa Severo Ochoa del Principado de Asturias (Spain); the Chulalongkorn Academic into Its 2nd Century Project Advancement Project, and the National Science, Research and Innovation Fund via the Program Management Unit for Human Resources \& Institutional Development, Research and Innovation, grant B37G660013 (Thailand); the Kavli Foundation; the Nvidia Corporation; the SuperMicro Corporation; the Welch Foundation, contract C-1845; and the Weston Havens Foundation (USA).
\end{acknowledgments}

\bibliography{auto_generated}

\providecommand{\href}[2]{#2}\begingroup\raggedright\begin{thebibliography}{10}%
\makeatletter
\providecommand{\hrefCMSnoop }[0]{\@secondoftwo}%
\makeatother
\providecommand{\doi}{\texttt{doi:}\begingroup \urlstyle{tt}\Url}

\bibitem{Shuryak:1977ut}
\hrefCMSnoop {}{E.~V. Shuryak, ``{Theory of hadronic plasma}'',} \textit{ Sov.
  Phys. JETP} \textbf{ 47} (1978)
212.

\bibitem{Collins:1974ky}
\hrefCMSnoop {}{J.~C. Collins and M.~J. Perry, ``{Superdense matter: neutrons
  or asymptotically free quarks?}'',} \textit{ Phys. Rev. Lett.} \textbf{ 34}
  (1975) 1353,
\href{http://dx.doi.org/10.1103/PhysRevLett.34.1353}{\doi{10.1103/PhysRevLett.34.1353}}.

\bibitem{Cabibbo:1975ig}
\hrefCMSnoop {}{N.~Cabibbo and G.~Parisi, ``{Exponential hadronic spectrum and
  quark liberation}'',} \textit{ Phys. Lett. B} \textbf{ 59} (1975) 67,
\href{http://dx.doi.org/10.1016/0370-2693(75)90158-6}{\doi{10.1016/0370-2693(75)90158-6}}.

\bibitem{Freedman:1976ub}
\hrefCMSnoop {}{B.~A. Freedman and L.~D. McLerran, ``{Fermions and gauge vector
  mesons at finite temperature and density. 3. the ground state energy of a
  relativistic quark gas}'',} \textit{ Phys. Rev. D} \textbf{ 16} (1977) 1169,
\href{http://dx.doi.org/10.1103/PhysRevD.16.1169}{\doi{10.1103/PhysRevD.16.1169}}.

\bibitem{Hwa:2004yg}
R.~C. Hwa and X.~N. Wang, eds., ``{Quark-gluon plasma 3}''.
\newblock World Scientific, Singapore, 2004.
\newblock \href{http://dx.doi.org/10.1142/5029}{\doi{10.1142/5029}},
  ISBN~978-981-238-077-7, 978-981-4488-08-2.

\bibitem{HotQCD:2014kol}
\hrefCMSnoop {}{{HotQCD} Collaboration, ``{Equation of state in (2+1)-flavor
  {QCD}}'',} \textit{ Phys. Rev. D} \textbf{ 90} (2014) 094503,
  \href{http://dx.doi.org/10.1103/PhysRevD.90.094503}{\doi{10.1103/PhysRevD.90.094503}},
  \href{http://www.arXiv.org/abs/1407.6387}{\texttt{arXiv:1407.6387}}.

\bibitem{Adams:2005dq}
\hrefCMSnoop {}{{STAR} Collaboration, ``Experimental and theoretical challenges
  in the search for the quark gluon plasma: The {STAR} collaboration's critical
  assessment of the evidence from {RHIC} collisions'',} \textit{ Nucl. Phys. A}
  \textbf{ 757} (2005) 102,
  \href{http://dx.doi.org/10.1016/j.nuclphysa.2005.03.085}{\doi{10.1016/j.nuclphysa.2005.03.085}},
  \href{http://www.arXiv.org/abs/nucl-ex/0501009}{\texttt{arXiv:nucl-ex/0501009}}.

\bibitem{Adcox:2004mh}
\hrefCMSnoop {}{{PHENIX} Collaboration, ``Formation of dense partonic matter in
  relativistic nucleus-nucleus collisions at {RHIC}: experimental evaluation by
  the {PHENIX} collaboration'',} \textit{ Nucl. Phys. A} \textbf{ 757} (2005)
  184,
  \href{http://dx.doi.org/10.1016/j.nuclphysa.2005.03.086}{\doi{10.1016/j.nuclphysa.2005.03.086}},
  \href{http://www.arXiv.org/abs/nucl-ex/0410003}{\texttt{arXiv:nucl-ex/0410003}}.

\bibitem{Arsene:2004fa}
\hrefCMSnoop {}{{BRAHMS} Collaboration, ``Quark gluon plasma and color glass
  condensate at {RHIC}? {T}he perspective from the {BRAHMS} experiment'',}
  \textit{ Nucl. Phys. A} \textbf{ 757} (2005) 1,
  \href{http://dx.doi.org/10.1016/j.nuclphysa.2005.02.130}{\doi{10.1016/j.nuclphysa.2005.02.130}},
  \href{http://www.arXiv.org/abs/nucl-ex/0410020}{\texttt{arXiv:nucl-ex/0410020}}.

\bibitem{Back:2004je}
\hrefCMSnoop {}{{PHOBOS} Collaboration, ``The {PHOBOS} perspective on
  discoveries at {RHIC}'',} \textit{ Nucl. Phys. A} \textbf{ 757} (2005) 28,
  \href{http://dx.doi.org/10.1016/j.nuclphysa.2005.03.084}{\doi{10.1016/j.nuclphysa.2005.03.084}},
  \href{http://www.arXiv.org/abs/nucl-ex/0410022}{\texttt{arXiv:nucl-ex/0410022}}.

\bibitem{Heinz:2013th}
\hrefCMSnoop {}{U.~Heinz and R.~Snellings, ``Collective flow and viscosity in
  relativistic heavy-ion collisions'',} \textit{ Ann. Rev. Nucl. Part. Sci.}
  \textbf{ 63} (2013) 123,
  \href{http://dx.doi.org/10.1146/annurev-nucl-102212-170540}{\doi{10.1146/annurev-nucl-102212-170540}},
  \href{http://www.arXiv.org/abs/1301.2826}{\texttt{arXiv:1301.2826}}.

\bibitem{2007PhRvL..98q0401J}
J.~{Joseph}\hrefCMSnoop {}{ { et~al.}, ``Measurement of sound velocity in a
  {F}ermi gas near a {F}eshbach resonance'',} \textit{ Phys. Rev. Lett.}
  \textbf{ 98} (2007) 170401,
  \href{http://dx.doi.org/10.1103/PhysRevLett.98.170401}{\doi{10.1103/PhysRevLett.98.170401}}.

\bibitem{2020Sci...370.1222P}
P.~B. {Patel}\hrefCMSnoop {}{ { et~al.}, ``{Universal sound diffusion in a
  strongly interacting Fermi gas}'',} \textit{ Science} \textbf{ 370} (2020)
  1222,
  \href{http://dx.doi.org/10.1126/science.aaz5756}{\doi{10.1126/science.aaz5756}},
  \href{http://www.arXiv.org/abs/1909.02555}{\texttt{arXiv:1909.02555}}.

\bibitem{Campanini:2011bj}
\hrefCMSnoop {}{R.~Campanini and G.~Ferri, ``{Experimental equation of state in
  proton-proton and proton-antiproton collisions and phase transition to quark
  gluon plasma}'',} \textit{ Phys. Lett. B} \textbf{ 703} (2011) 237,
  \href{http://dx.doi.org/10.1016/j.physletb.2011.08.009}{\doi{10.1016/j.physletb.2011.08.009}},
  \href{http://www.arXiv.org/abs/1106.2008}{\texttt{arXiv:1106.2008}}.

\bibitem{Pratt:2015zsa}
\hrefCMSnoop {}{S.~Pratt, E.~Sangaline, P.~Sorensen, and H.~Wang,
  ``{Constraining the equation of state of super-hadronic matter from heavy-ion
  collisions}'',} \textit{ Phys. Rev. Lett.} \textbf{ 114} (2015) 202301,
  \href{http://dx.doi.org/10.1103/PhysRevLett.114.202301}{\doi{10.1103/PhysRevLett.114.202301}},
  \href{http://www.arXiv.org/abs/1501.04042}{\texttt{arXiv:1501.04042}}.

\bibitem{Gardim:2019xjs}
\hrefCMSnoop {}{F.~G. Gardim, G.~Giacalone, M.~Luzum, and J.-Y. Ollitrault,
  ``{Thermodynamics of hot strong-interaction matter from ultrarelativistic
  nuclear collisions}'',} \textit{ Nature Phys.} \textbf{ 16} (2020) 615,
  \href{http://dx.doi.org/10.1038/s41567-020-0846-4}{\doi{10.1038/s41567-020-0846-4}},
  \href{http://www.arXiv.org/abs/1908.09728}{\texttt{arXiv:1908.09728}}.

\bibitem{Gardim:2019brr}
\hrefCMSnoop {}{F.~G. Gardim, G.~Giacalone, and J.-Y. Ollitrault, ``{The mean
  transverse momentum of ultracentral heavy-ion collisions: A new probe of
  hydrodynamics}'',} \textit{ Phys. Lett. B} \textbf{ 809} (2020) 135749,
  \href{http://dx.doi.org/10.1016/j.physletb.2020.135749}{\doi{10.1016/j.physletb.2020.135749}},
  \href{http://www.arXiv.org/abs/1909.11609}{\texttt{arXiv:1909.11609}}.

\bibitem{Sorensen:2021zme}
\hrefCMSnoop {}{A.~Sorensen, D.~Oliinychenko, V.~Koch, and L.~McLerran, ``Speed
  of sound and baryon cumulants in heavy-ion collisions'',} \textit{ Phys. Rev.
  Lett.} \textbf{ 127} (2021) 042303,
  \href{http://dx.doi.org/10.1103/PhysRevLett.127.042303}{\doi{10.1103/PhysRevLett.127.042303}},
  \href{http://www.arXiv.org/abs/2103.07365}{\texttt{arXiv:2103.07365}}.

\bibitem{Nijs:2021clz}
\hrefCMSnoop {}{G.~Nijs and W.~van~der Schee, ``{Predictions and postdictions
  for relativistic lead and oxygen collisions with the computational simulation
  code Trajectum}'',} \textit{ Phys. Rev. C} \textbf{ 106} (2022) 044903,
  \href{http://dx.doi.org/10.1103/PhysRevC.106.044903}{\doi{10.1103/PhysRevC.106.044903}},
  \href{http://www.arXiv.org/abs/2110.13153}{\texttt{arXiv:2110.13153}}.

\bibitem{Qiu:2012tm}
\hrefCMSnoop {}{Z.~Qiu, C.~Shen, and U.~W. Heinz, ``{Resonance decay
  contributions to higher-order anisotropic flow coefficients}'',} \textit{
  Phys. Rev. C} \textbf{ 86} (2012) 064906,
  \href{http://dx.doi.org/10.1103/PhysRevC.86.064906}{\doi{10.1103/PhysRevC.86.064906}},
  \href{http://www.arXiv.org/abs/1210.7010}{\texttt{arXiv:1210.7010}}.

\bibitem{hepdata}
\hrefCMSnoop {}{{CMS Collaboration}, ``{{HEPD}ata record for this analysis}'',}
  2023.
\newblock
  \href{http://dx.doi.org/10.17182/hepdata.146016}{\doi{10.17182/hepdata.146016}}.

\bibitem{CMS:2008xjf}
\hrefCMSnoop {}{{CMS Collaboration}, ``The {CMS} experiment at the {CERN}
  {LHC}'',} \textit{ JINST} \textbf{ 3} (2008) S08004,
  \href{http://dx.doi.org/10.1088/1748-0221/3/08/S08004}{\doi{10.1088/1748-0221/3/08/S08004}}.

\bibitem{CMS:2020cmk}
\hrefCMSnoop {}{{CMS Collaboration}, ``{Performance of the CMS level-1 trigger
  in proton-proton collisions at $\sqrt{s} = 13$\TeV}'',} \textit{ JINST}
  \textbf{ 15} (2020) P10017,
  \href{http://dx.doi.org/10.1088/1748-0221/15/10/P10017}{\doi{10.1088/1748-0221/15/10/P10017}},
  \href{http://www.arXiv.org/abs/2006.10165}{\texttt{arXiv:2006.10165}}.

\bibitem{Khachatryan:2016bia}
\hrefCMSnoop {}{{CMS Collaboration}, ``The {CMS} trigger system'',} \textit{
  JINST} \textbf{ 12} (2017) P01020,
  \href{http://dx.doi.org/10.1088/1748-0221/12/01/P01020}{\doi{10.1088/1748-0221/12/01/P01020}},
  \href{http://www.arXiv.org/abs/1609.02366}{\texttt{arXiv:1609.02366}}.

\bibitem{CMS:2020uim}
\hrefCMSnoop {}{{CMS Collaboration}, ``Electron and photon reconstruction and
  identification with the {CMS} experiment at the {CERN} {LHC}'',} \textit{
  JINST} \textbf{ 16} (2021) P05014,
  \href{http://dx.doi.org/10.1088/1748-0221/16/05/P05014}{\doi{10.1088/1748-0221/16/05/P05014}},
  \href{http://www.arXiv.org/abs/2012.06888}{\texttt{arXiv:2012.06888}}.

\bibitem{CMS:2018rym}
\hrefCMSnoop {}{{CMS Collaboration}, ``Performance of the {CMS} muon detector
  and muon reconstruction with proton-proton collisions at
  $\sqrt{s}=13${\TeV}'',} \textit{ JINST} \textbf{ 13} (2018) P06015,
  \href{http://dx.doi.org/10.1088/1748-0221/13/06/P06015}{\doi{10.1088/1748-0221/13/06/P06015}},
  \href{http://www.arXiv.org/abs/1804.04528}{\texttt{arXiv:1804.04528}}.

\bibitem{CMS:2014pgm}
\hrefCMSnoop {}{{CMS Collaboration}, ``{Description and performance of track
  and primary-vertex reconstruction with the CMS tracker}'',} \textit{ JINST}
  \textbf{ 9} (2014) P10009,
  \href{http://dx.doi.org/10.1088/1748-0221/9/10/P10009}{\doi{10.1088/1748-0221/9/10/P10009}},
  \href{http://www.arXiv.org/abs/1405.6569}{\texttt{arXiv:1405.6569}}.

\bibitem{CMS:2017yfk}
\hrefCMSnoop {}{{CMS Collaboration}, ``Particle-flow reconstruction and global
  event description with the {CMS} detector'',} \textit{ JINST} \textbf{ 12}
  (2017) P10003,
  \href{http://dx.doi.org/10.1088/1748-0221/12/10/P10003}{\doi{10.1088/1748-0221/12/10/P10003}},
\href{http://www.arXiv.org/abs/1706.04965}{\texttt{arXiv:1706.04965}}.

\bibitem{Bayatian:2006jz}
\hrefCMSnoop {}{G.~Bayatian { et~al.}, ``{Design, performance and calibration
  of the CMS forward calorimeter wedges}'',} \textit{ Eur. Phys. J. C} \textbf{
  53} (2008) 139,
  \href{http://dx.doi.org/10.1140/epjc/s10052-007-0459-4}{\doi{10.1140/epjc/s10052-007-0459-4}}.

\bibitem{Suranyi:2021ssd}
\hrefCMSnoop {}{O.~Sur{\'a}nyi { et~al.}, ``Performance of the {CMS} zero
  degree calorimeters in {pPb} collisions at the {LHC}'',} \textit{ JINST}
  \textbf{ 16} (2021) P05008,
  \href{http://dx.doi.org/10.1088/1748-0221/16/05/P05008}{\doi{10.1088/1748-0221/16/05/P05008}},
  \href{http://www.arXiv.org/abs/2102.06640}{\texttt{arXiv:2102.06640}}.

\bibitem{CMS:2016xef}
\hrefCMSnoop {}{{CMS Collaboration}, ``{Charged-particle nuclear modification
  factors in PbPb and pPb collisions at $\sqrtsNN = 5.02\TeV$}'',} \textit{
  JHEP} \textbf{ 04} (2017) 039,
  \href{http://dx.doi.org/10.1007/JHEP04(2017)039}{\doi{10.1007/JHEP04(2017)039}},
  \href{http://www.arXiv.org/abs/1611.01664}{\texttt{arXiv:1611.01664}}.

\bibitem{CMS:2013bza}
\hrefCMSnoop {}{{CMS Collaboration}, ``{Studies of azimuthal dihadron
  correlations in ultra-central \PbPb collisions at $\sqrtsNN = 2.76\TeV$}'',}
  \textit{ JHEP} \textbf{ 02} (2014) 088,
  \href{http://dx.doi.org/10.1007/JHEP02(2014)088}{\doi{10.1007/JHEP02(2014)088}},
  \href{http://www.arXiv.org/abs/1312.1845}{\texttt{arXiv:1312.1845}}.

\bibitem{CMS:2019nct}
\hrefCMSnoop {}{{CMS Collaboration}, ``{Mixed higher-order anisotropic flow and
  nonlinear response coefficients of charged particles in \PbPb collisions at
  $\sqrtsNN = 2.76$ and 5.02\TeV}'',} \textit{ Eur. Phys. J. C} \textbf{ 80}
  (2020) 534,
  \href{http://dx.doi.org/10.1140/epjc/s10052-020-7834-9}{\doi{10.1140/epjc/s10052-020-7834-9}},
  \href{http://www.arXiv.org/abs/1910.08789}{\texttt{arXiv:1910.08789}}.

\bibitem{Lokhtin:2008xi}
I.~P. Lokhtin\hrefCMSnoop {}{ { et~al.}, ``{Heavy ion event generator HYDJET++
  (HYDrodynamics plus JETs)}'',} \textit{ Comput. Phys. Commun.} \textbf{ 180}
  (2009) 779,
  \href{http://dx.doi.org/10.1016/j.cpc.2008.11.015}{\doi{10.1016/j.cpc.2008.11.015}},
  \href{http://www.arXiv.org/abs/0809.2708}{\texttt{arXiv:0809.2708}}.

\bibitem{Agostinelli:2002hh}
\hrefCMSnoop {}{{GEANT4} Collaboration, ``{GEANT4---a simulation toolkit}'',}
  \textit{ Nucl. Instrum. Meth. A} \textbf{ 506} (2003) 250,
  \href{http://dx.doi.org/10.1016/S0168-9002(03)01368-8}{\doi{10.1016/S0168-9002(03)01368-8}}.

\bibitem{Hagedorn:1983wk}
\hrefCMSnoop {}{R.~Hagedorn, ``{Multiplicities, \pt distributions and the
  expected hadron to quark-gluon phase transition}'',} \textit{ Riv. Nuovo
  Cim.} \textbf{ 6N10} (1983) 1,
  \href{http://dx.doi.org/10.1007/BF02740917}{\doi{10.1007/BF02740917}}.

\bibitem{ALICE:2018hza}
\hrefCMSnoop {}{{ALICE Collaboration}, ``{Transverse momentum spectra and
  nuclear modification factors of charged particles in Xe-Xe collisions at
  $\sqrtsNN = 5.44\TeV$}'',} \textit{ Phys. Lett. B} \textbf{ 788} (2019) 166,
  \href{http://dx.doi.org/10.1016/j.physletb.2018.10.052}{\doi{10.1016/j.physletb.2018.10.052}},
  \href{http://www.arXiv.org/abs/1805.04399}{\texttt{arXiv:1805.04399}}.

\bibitem{ALICE:2015juo}
\hrefCMSnoop {}{{ALICE Collaboration}, ``{Centrality dependence of the
  charged-particle multiplicity density at midrapidity in Pb-Pb collisions at
  $\sqrtsNN = 5.02\TeV$}'',} \textit{ Phys. Rev. Lett.} \textbf{ 116} (2016)
  222302,
  \href{http://dx.doi.org/10.1103/PhysRevLett.116.222302}{\doi{10.1103/PhysRevLett.116.222302}},
  \href{http://www.arXiv.org/abs/1512.06104}{\texttt{arXiv:1512.06104}}.

\bibitem{Das:2017ned}
\hrefCMSnoop {}{S.~J. Das, G.~Giacalone, P.-A. Monard, and J.-Y. Ollitrault,
  ``{Relating centrality to impact parameter in nucleus-nucleus collisions}'',}
  \textit{ Phys. Rev. C} \textbf{ 97} (2018) 014905,
  \href{http://dx.doi.org/10.1103/PhysRevC.97.014905}{\doi{10.1103/PhysRevC.97.014905}},
  \href{http://www.arXiv.org/abs/1708.00081}{\texttt{arXiv:1708.00081}}.

\bibitem{Giacalone:2023cet}
\hrefCMSnoop {}{G.~Giacalone, G.~Nijs, and W.~van~der Schee, ``{Determination
  of the neutron skin of Pb208 from ultrarelativistic nuclear collisions}'',}
  \textit{ Phys. Rev. Lett.} \textbf{ 131} (2023) 202302,
  \href{http://dx.doi.org/10.1103/PhysRevLett.131.202302}{\doi{10.1103/PhysRevLett.131.202302}},
  \href{http://www.arXiv.org/abs/2305.00015}{\texttt{arXiv:2305.00015}}.

\bibitem{Nijs:2023bzv}
\hrefCMSnoop {}{G.~Nijs and W.~van~der Schee, ``{Ultracentral heavy ion
  collisions, transverse momentum and the equation of state}'',} 2023.
  \href{http://www.arXiv.org/abs/2312.04623}{\texttt{arXiv:2312.04623}}.

\bibitem{Borsanyi:2010cj}
S.~Borsanyi\hrefCMSnoop {}{ { et~al.}, ``{The QCD equation of state with
  dynamical quarks}'',} \textit{ JHEP} \textbf{ 11} (2010) 077,
  \href{http://dx.doi.org/10.1007/JHEP11(2010)077}{\doi{10.1007/JHEP11(2010)077}},
  \href{http://www.arXiv.org/abs/1007.2580}{\texttt{arXiv:1007.2580}}.

\end{thebibliography}\endgroup
\cleardoublepage \appendix\section{The CMS Collaboration \label{app:collab}}\begin{sloppypar}\hyphenpenalty=5000\widowpenalty=500\clubpenalty=5000
\cmsinstitute{Yerevan Physics Institute, Yerevan, Armenia}
{\tolerance=6000
A.~Hayrapetyan, A.~Tumasyan\cmsAuthorMark{1}\cmsorcid{0009-0000-0684-6742}
\par}
\cmsinstitute{Institut f\"{u}r Hochenergiephysik, Vienna, Austria}
{\tolerance=6000
W.~Adam\cmsorcid{0000-0001-9099-4341}, J.W.~Andrejkovic, T.~Bergauer\cmsorcid{0000-0002-5786-0293}, S.~Chatterjee\cmsorcid{0000-0003-2660-0349}, K.~Damanakis\cmsorcid{0000-0001-5389-2872}, M.~Dragicevic\cmsorcid{0000-0003-1967-6783}, P.S.~Hussain\cmsorcid{0000-0002-4825-5278}, M.~Jeitler\cmsAuthorMark{2}\cmsorcid{0000-0002-5141-9560}, N.~Krammer\cmsorcid{0000-0002-0548-0985}, A.~Li\cmsorcid{0000-0002-4547-116X}, D.~Liko\cmsorcid{0000-0002-3380-473X}, I.~Mikulec\cmsorcid{0000-0003-0385-2746}, J.~Schieck\cmsAuthorMark{2}\cmsorcid{0000-0002-1058-8093}, R.~Sch\"{o}fbeck\cmsorcid{0000-0002-2332-8784}, D.~Schwarz\cmsorcid{0000-0002-3821-7331}, M.~Sonawane\cmsorcid{0000-0003-0510-7010}, S.~Templ\cmsorcid{0000-0003-3137-5692}, W.~Waltenberger\cmsorcid{0000-0002-6215-7228}, C.-E.~Wulz\cmsAuthorMark{2}\cmsorcid{0000-0001-9226-5812}
% ...
\par}
\vskip\cmsinstskip
\dag:~Deceased\\
$^{1}$Also at Yerevan State University, Yerevan, Armenia\\
$^{2}$Also at TU Wien, Vienna, Austria\\
% ...
\end{sloppypar}
\end{document}